\documentclass[twocolumn,prb,superscriptaddress,showpacs,amsmath,amssymb]{revtex4}
\usepackage{bm,graphicx,dcolumn}
\newcommand{\pdt}{\partial_{t}}

\begin{document}
\title{Shapiro steps as a direct probe of $\pm s$-wave symmetry in
multigap superconducting Josephson junctions}

\affiliation{
CCSE, Japan Atomic Energy Agency, 
6-9-3 Higashi-Ueno Taito-ku, Tokyo 110-0015, Japan}
\affiliation{
Institute for Materials Research, Tohoku University, 2-1-1 Katahira
Aoba-ku, Sendai 980-8577, Japan} 
\affiliation{
CREST(JST), 4-1-8 Honcho, Kawaguchi, Saitama 332-0012, Japan}
\affiliation{
JST, TRIP, 5 Sambancho Chiyoda-ku, Tokyo 102-0075, Japan}
\author{Yukihiro Ota}
\affiliation{
CCSE, Japan Atomic Energy Agency, 
6-9-3 Higashi-Ueno Taito-ku, Tokyo 110-0015, Japan}
\affiliation{
CREST(JST), 4-1-8 Honcho, Kawaguchi, Saitama 332-0012, Japan}
\author{Masahiko Machida}
\affiliation{
CCSE, Japan Atomic Energy Agency, 
6-9-3 Higashi-Ueno Taito-ku, Tokyo 110-0015, Japan}
\affiliation{
CREST(JST), 4-1-8 Honcho, Kawaguchi, Saitama 332-0012, Japan}
\affiliation{
JST, TRIP, 5 Sambancho Chiyoda-ku, Tokyo 102-0075, Japan}
\author{Tomio Koyama}
\affiliation{
Institute for Materials Research, Tohoku University, 
2-1-1 Katahira Aoba-ku, Sendai 980-8577, Japan}
\affiliation{
CREST(JST), 4-1-8 Honcho, Kawaguchi, Saitama 332-0012, Japan}
\date{\today}

\begin{abstract}
We theoretically study the Shapiro steps in a hetero-Josephson junction
 made of a single-gap superconductor and a two-gap one. 
We find that an anomalous dc Josephson current is induced by tuning the
 frequency of an applied microwave to the Josephson-Leggett mode
 frequency, which creates an extra step structure in the $I$-$V$
 characteristics besides the conventional Shapiro steps. 
The step heights at the resonance voltages exhibit an alternate
 structure of a large and small value reflecting the gap symmetry of the
 two-gap superconductor. 
In the $\pm s$-wave case in which the two gaps have opposite signs in
 the two-gap superconductor the steps with odd index are enhanced,
 whereas in the s-wave case the ones with even index have larger values.
The existence of the fractional Shapiro steps is also predicted. 
\end{abstract}

\pacs{74.50.+r,74.70.Xa}
\maketitle

Since the discovery of iron-pnictide 
superconductors\,\cite{Kamihara;Hosono:2008,Rotter;Johrendt:2008}, the
symmetry of the superconducting gaps has attracted a great interest. 
The spin-fluctuation mechanism based on nesting between disconnected 
multiple Fermi surfaces predicts the $\pm s$-wave
symmetry\,\cite{Mazin;Du:2008,Kuroki;Aoki:2008}.  
It has been reported that the scanning tunneling microscope exhibits a
characteristic field dependence expected from the $\pm s$-wave
symmetry\,\cite{Hanaguri;Takagi:2010}, and also  
the scanning SQUID detects half-fluxon dynamics consistent with 
the $\pm s$-wave gap\,\cite{Chen;Zha:2010}. 
However, the consensus on the pairing symmetry of the iron-based 
superconductors have not yet been attained\,\cite{Li;Zhang:2010}.    
Hence, a more definitive experimental probe is now in great demand. 

To clarify the pairing symmetry and associated new physics, various
studies have been done for the Josephson effects in the iron-based
superconductors both
experimentally\,\cite{Zhang;Takeuchi:2009,Wu;Wu:2010,Kashiwaya;Kashiwaya:2010}
and
theoretically\,\cite{Parker;Mazin:2009,Wu;Phillips:2009,Ng;Nagaosa:2009,Golubov;Dolgov:2009,Linder;Sudbo:2009,Inotani;Ohashi:2009,Ota;Matsumoto:2009}. 
The present authors\,\cite{Ota;Matsumoto:2009}
have shown theoretically that a
superconducting-insulator-superconducting (SIS) Josephson junction made
of single-gap and multi-gap superconductors reveals Josephson
effects depending sensitively on the gap symmetry.  
In such a hetero Josephson junction the Cooper-pairs can transfer via
multiple tunneling channels between the superconducting electrodes, because
one of the electrodes has multiple conduction bands available for
the tunneling.     
In the case of the $\pm s$-wave symmetry the two tunneling channels,
i.e., ``0-junction'' and ``$\pi$-junction'' are formed.  
Therefore, one expects unusual Josephson effects different
from the conventional one. 
Furthermore, a new phase oscillation mode called the ``Josephson-Leggett
(JL) mode''\,\cite{Ota;Matsumoto:2009,Leggett:1966} 
exists in the low frequency region besides the Josephson plasma. 
This mode is expected to affect the AC Josephson effect. 
In this paper, we investigate effects of the JL mode
on Shapiro steps\,\cite{Hinken:1991} in such an SIS hetero Josephson
junction under microwave irradiation. 
We show that resonance between the JL mode and an applied microwave
occurs in the phase running states when the microwave frequency is tuned to the
frequency of the JL mode. 
The maximum DC Josephson currents induced at the resonant voltages exhibit
remarkable behavior reflecting the gap symmetry.  
Furthermore, the existence of fractional Shapiro steps is predicted. 

\begin{figure}[tbp]
\centering
\scalebox{0.3}[0.3]{\includegraphics{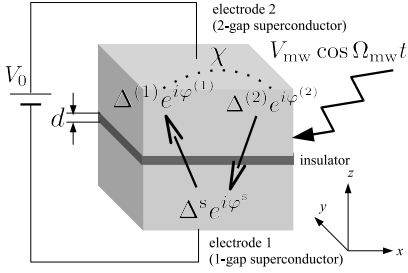}}
\caption{Schematic figure for a hetero Josephson junction made of
single-gap and two-gap superconductors. 
A microwave, whose amplitude and frequency are, respectively, 
$V_{\rm mw}$ and $\Omega_{\rm mw}$, is applied
 on the voltage-biased Josephson junction. } 
\label{fig:acheteroJ}
\end{figure}

Consider an SIS hetero Josephson junction shown schematically in
Fig.\,\ref{fig:acheteroJ}.  
We assume that the electrodes $1$ and $2$ in this system are,
respectively, single- and two-gap superconductors and the insulating
barrier between the two electrodes has width $d$ and dielectric constant $\epsilon$.  
To observe the Shapiro steps we irradiate a microwave generating the AC
voltage $V_{\rm mw}\cos\Omega_{\rm mw} t$ at the junction site as shown
in Fig.\,\ref{fig:acheteroJ}.

Assuming that the phase difference is uniform along the in-plane
direction ($\parallel xy$-plane), one can derive  
the effective Lagrangian density in the present hetero Josephson junction system
under no external magnetic field\cite{Ota;Matsumoto:2009} as follows 
\begin{eqnarray}
&&
 \mathcal{L}_{\rm eff}
=
\frac{q_{\rm s}^{2}}{8\pi \mu^{\prime}}
+
\sum_{i=1}^{2}\frac{ q_{i}^{2}}{8\pi \mu_{i}}
+
\sum_{i=1}^{2}\frac{\hbar j_{i}}{e^{\ast}}\cos\theta^{(i)} 
\nonumber \\
&&
\qquad\quad
+
\frac{\hbar J_{\rm in}}{e^{\ast}}\cos\psi
+
\frac{d\epsilon}{8\pi} (E_{21}^{z})^{2},
\label{eq:eff_lag}
\end{eqnarray}
where $\theta^{(i)}$ and $\psi$ are, respectively,  the gauge-invariant
phase difference and the relative phase of the superconducting gaps
having the phases $\varphi^{(1)}$ and $\varphi^{(2)}$ in the electrode
$2$ defined as 
\begin{equation*}
\theta^{(i)} 
= 
\varphi^{(i)} - \varphi^{\rm s} - \frac{e^{\ast}d}{\hbar c}A_{21}^{z},
\qquad
\psi
=
\varphi^{(1)}-\varphi^{(2)},
\end{equation*} 
with $\varphi^{(s)}$ being the phase of the electrode $1$. 
The quantity $A_{21}^z$  in $\theta^{(i)}$ is given as 
\(
A_{21}^{z} = d^{-1}\int_{-d/2}^{d/2}A^{z}(z) dz
\), 
$A^{z}(z)$ being the $z$ component of the vector potential. 
The first and the second terms in Eq.\,(\ref{eq:eff_lag})
represent the energy originating from the finite charge compressibility
inside the superconducting electrodes where the compressible area is
confined around the interface whose thickness is 
characterized by the charge screening length $\mu^{\prime}$ (for the
electrode $1$) or $\mu_{i}$ (for the electrode $2$). 
We define $q_{\rm s}$ and $q_{i}$ as, respectively, 
\(
q_{\rm s} = (\hbar/e^{\ast})\pdt\varphi^{\rm s} + A_{1}^{0}
\) and 
\(
q_{i} = (\hbar/e^{\ast})\pdt\varphi^{(i)} + A_{2}^{0}
\) with $e^{\ast}=2e$ and $A_{\ell}^{0}$ being the scalar potential 
in the electrode $\ell(=1,2)$.
The third term in Eq.\,(\ref{eq:eff_lag}) is the Josephson coupling
energy in the two tunneling channels with critical current values
$j_{1}$ and $j_2$. 
The forth term in Eq.\,(\ref{eq:eff_lag}) is the inter-band Josephson
coupling energy, which originates from the Cooper-pair transfer between
the two bands in the electrode $2$. 
The coupling constant $J_{\rm in}$ can take both signs $+$ and $-$. 
In the case of the $\pm s$-wave ($s$-wave) we have $J_{\rm in}<0$
($J_{\rm in}>0$). 
Note that the inter-band coupling energy is minimum at $\psi=\pi$ if
$J_{\rm in}<0$.  
The fifth term in Eq.\,(\ref{eq:eff_lag}) is the energy of the electric
field.  
The electric field inside the insulating layer is given as 
\(
 E_{21}^{z} 
= -c^{-1}\pdt A_{21}^{z} - d^{-1}(A^{0}_{2}-A^{0}_{1})
\). 

\begin{figure}[tbp]
\centering
\scalebox{0.54}[0.54]{\includegraphics{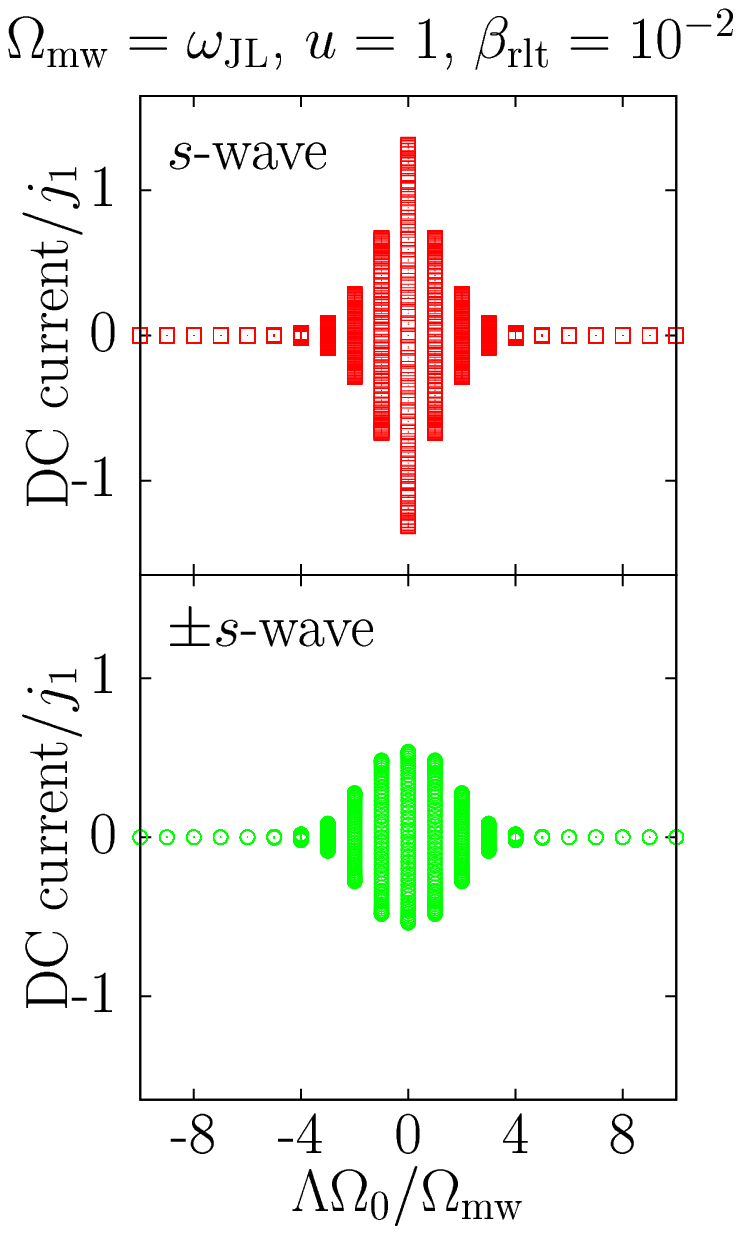}}
\scalebox{0.54}[0.54]{\includegraphics{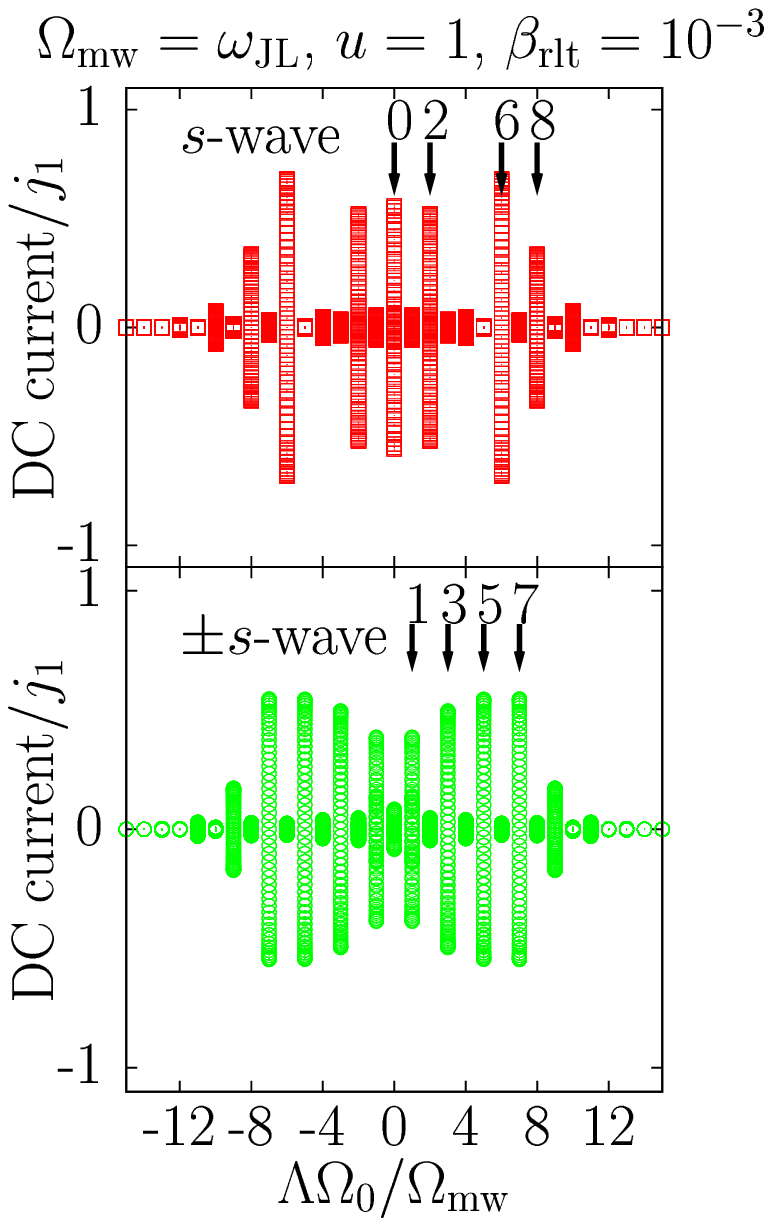}}
\vspace{-7mm}
\begin{flushleft}
(a)\hspace{38mm}(b)
\end{flushleft}
\vspace{-3mm}
\caption{DC components of $I_{\rm s}(t)$
 vs. $\Omega_{0}(=e^{\ast}V_{0}/\hbar)$. The applied microwave has the frequency  
$\Omega_{\rm mw}=\omega_{\rm JL}$ (i.e., $q=1$) and the amplitude
$u(=e^{\ast}V_{\rm mw}/\hbar\Omega_{\rm mw})=1$. 
The junction parameters are chosen as $j_{2}/j_{1}=0.93$,
 $\alpha_{2}/\alpha_{1}=1.3$, and $\alpha_{1}=0.1$.  
(a) $\beta_{\rm rlt}=10^{-2}$ and (b) $\beta_{\rm rlt}=10^{-3}$. 
In (b), the steps labeled by even (odd) numbers are larger in the $s$-wave ( $\pm s$-wave) case.}
\label{fig:q1} 
\end{figure}

Let us study the dynamics of $\theta^{(i)}$. From the effective
Lagrangian density (\ref{eq:eff_lag}) it follows the Josephson
relation\,\cite{Ota;Matsumoto:2009}  
\begin{equation*}
\sum_{i=1}^{2}\frac{\bar{\alpha}}{\alpha_{i}} \pdt\theta^{(i)} 
= \Lambda \frac{e^{\ast}d}{\hbar}E_{21}^{z},
\end{equation*}
where 
\(
\alpha^{\prime} = \epsilon \mu^{\prime }/d
\),
\(
\alpha_{i} = \epsilon \mu_{i}/d
\), 
$\bar{\alpha}^{-1} = \alpha_{1}^{-1} + \alpha_{2}^{-1}$ and
$\Lambda=1+\alpha^{\prime}+\bar{\alpha}$. 
When the voltage appearing in the junction is given as 
\(
dE_{21}^{z} = V_{0} + V_{\rm mw}\cos\Omega_{\rm mw} t
\),  we obtain   
\begin{subequations}
\begin{eqnarray}
\theta^{(1)}(t) 
&=& 
\theta_{0}^{(1)} + \Lambda f(t) 
+ \frac{\alpha_{1}}{\alpha_{1}+\alpha_{2}}\widetilde{\psi}(t), 
\label{eq:theta_one}\\
\theta^{(2)}(t)
&=& 
\theta_{0}^{(2)} + \Lambda f(t)
- \frac{\alpha_{2}}{\alpha_{1}+\alpha_{2}}\widetilde{\psi}(t), 
\label{eq:theta_two}
\end{eqnarray}
\end{subequations} 
where
\(
f(t) = \Omega_{0}t + u\sin\Omega_{\rm mw} t
\), 
$\Omega_{0} = e^{\ast}V_{0}/\hbar$, and 
$u=e^{\ast}V_{\rm mw}/\hbar \Omega_{\rm mw}$. 
The phases $\theta_{0}^{(1)}$ and $\theta_{0}^{(2)}$ are constants satisfying the relation,
$\psi_{0}=\theta_{0}^{(1)}-\theta_{0}^{(2)}=0\ (\pi$) for $J_{\rm in}>0$ ($<0$).  
The quantity $\widetilde{\psi}(t)$ describing the relative phase fluctuation,  
\(
 \widetilde{\psi}(t) = \psi(t) - \psi_{0}
\).
In this paper $\widetilde{\psi}$ is assumed to be small. 
Since the total Josephson current in the present system is given as  
\begin{equation}
 I_{\rm s}(t) = 
\sum_{i=1}^{2}j_{i}\sin [\theta^{(i)}(t)], 
\label{eq:sc}
\end{equation}
we have a simple expression, 
\(
I_{\rm s} 
= j_{\rm c}(\psi_{0})\sin[\theta_{0}^{(1)}+\Lambda f(t)]
\), for  $\widetilde{\psi}(t)=0$ with 
\(
j_{\rm c}(\psi_{0}) = j_{1} + j_{2}\cos\psi_{0}
\). 
Then, using the standard technique\,\cite{Hinken:1991}, 
one finds that the Shapiro steps appear at the voltages which are the
same as in the conventional Josephson junctions, i.e.,     
\(
V_{0} = -n\Lambda ^{-1}(\hbar/e^{\ast})\Omega_{\rm mw} 
\), $n$ being an integer. 
We note that the step height 
\(
|j_{\rm c}(\psi_{0})J_{n}(\Lambda u)|
\) depends on the relative phase, that is, the value of the critical
current at each step is enhanced (suppressed) for the $s$-wave 
($\pm s$-wave) case.    
The critical current depending on the gap-symmetry also leads
to the anomaly of the Reidel peak, which can be used for the
identification of the $\pm s$-wave gap as proposed by Inotani and
Ohashi\,\cite{Inotani;Ohashi:2009}. 

\begin{figure}[tbp]
\centering
\scalebox{0.52}[0.52]{\includegraphics{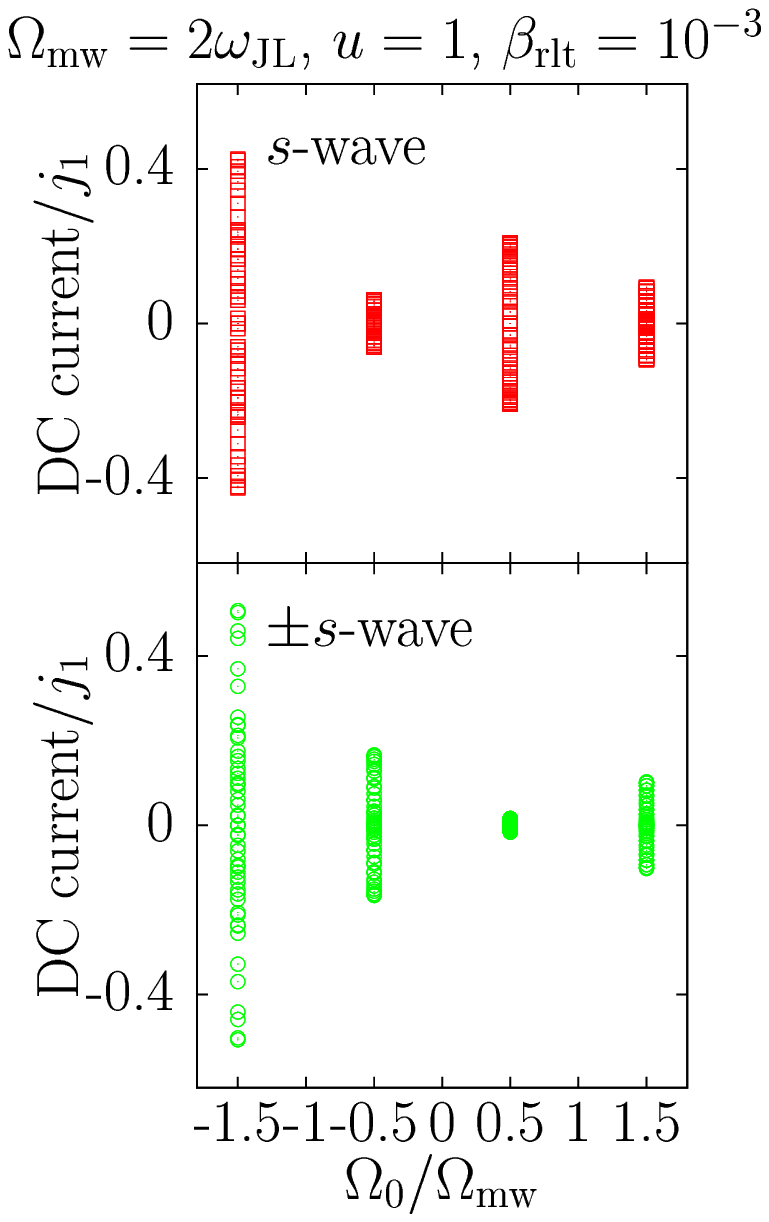}}
\scalebox{0.52}[0.52]{\includegraphics{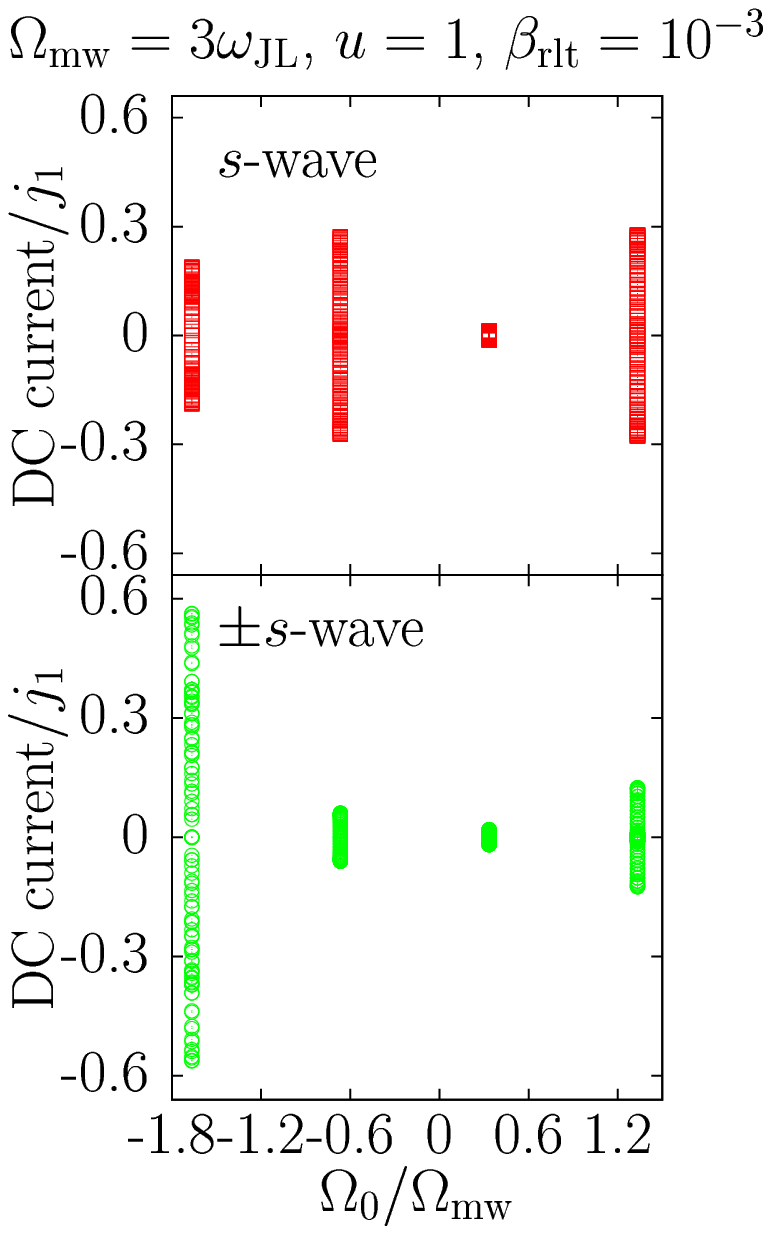}}
\vspace{-7mm}
\begin{flushleft}
(a)\hspace{35mm}(b)
\end{flushleft}
\vspace{-3mm} 
\caption{DC components of $I_{\rm s}(t)$
 vs. $\Omega_{0}(=e^{\ast}V_{0}/\hbar)$. 
The frequency and the amplitude of an applied microwave, respectively, 
$\Omega_{\rm mw}=q\omega_{\rm JL}$ ($q\neq 1$) and 
$u(=e^{\ast}V_{\rm mw}/\hbar\Omega_{\rm mw})=1$. 
The detuning $\beta_{\rm rlt}=10^{-3}$. 
(a) $q=2$ (b) $q=3$. }
\label{fig:qfrac} 
\end{figure}

Next suppose that $\widetilde{\psi}(t)$ is temporally fluctuated. 
To describe the time evolution of $\widetilde{\psi}$ we utilize the
equation of motion obtained from Eq.\,(\ref{eq:eff_lag}).  
From the  Euler-Lagrange equations with respect to $\varphi^{\rm s}$ 
and $\varphi^{(i)}$ one can derive the equation as 
\,\cite{Ota;Matsumoto:2009}, 
\begin{eqnarray}
&&
\frac{1}{\alpha_{1}}\pdt^{2}\theta^{(1)}
-
\frac{1}{\alpha_{2}}\pdt^{2}\theta^{(2)}
+ 
\xi \frac{1+\alpha^{\prime}}{\Lambda} 
\sum_{i=1}^{2}\frac{1}{\alpha_{i}} \pdt^{2}\theta^{(i)} 
\nonumber \\
&=&
-\omega_{1}^{2} \sin\theta^{(1)} + \omega_{2}^{2}\sin\theta^{(2)}
-{\rm sgn}(J_{\rm in}) 2\omega_{\rm in}^{2} \sin\psi,
\label{eq:rlt_eq} 
\end{eqnarray}
where 
\(
\omega_{i} = \sqrt{4\pi e^{\ast}dj_{i}/\epsilon \hbar}
\),
\(
\omega_{\rm in} = \sqrt{4\pi e^{\ast}d|J_{\rm in}|/\epsilon \hbar}
\), and 
\(
\xi = (\alpha_{1}-\alpha_{2})/(\alpha_{1}+\alpha_{2})
\). 
We note that Eq.\,(\ref{eq:rlt_eq}) describes the effect that the charge
imbalance between the two  bands in the electrode $2$ (the
left-hand-side of the equation) is induced by the Josephson current
between the two electrodes (the right-hand-side). 
The charge imbalance in this system is characterized by the parameter
$\xi$. 
Substituting Eqs.\,(\ref{eq:theta_one}) and (\ref{eq:theta_two}) into
Eq.\,(\ref{eq:rlt_eq}), we find the linearized equation for small
$\widetilde{\psi}$ , which is similar to the one for a forced harmonic
oscillator, 
\begin{equation}
 \pdt^{2}\widetilde{\psi} 
= 
-[\omega_{\rm JL}^{2} + G_{\rm J}(t)]\widetilde{\psi} 
+ F_{\rm C}(t) + F_{\rm J}(t),
\label{eq:rp_fluctuation}
\end{equation}
where 
\begin{eqnarray*}
F_{\rm C }(t)
&=& 
-\xi \frac{\alpha_{1}+\alpha_{2}}{2}\Omega_{\rm mw}^{2}
u\sin\Omega_{\rm mw} t, \\
F_{\rm J}(t)
&=&
-\nu \frac{\alpha_{1}+\alpha_{2}}{2}(\omega_{1}^{2}+\omega_{2}^{2})
\sin[\theta^{(1)}_{0}+\Lambda f(t)], 
\end{eqnarray*} 
with 
\(
\nu = [\omega_{1}^{2}-{\rm sgn}(J_{\rm in})\omega_{2}^{2}]
/(\omega_{1}^{2}+\omega_{2}^{2})
\)
and 
\(
G_{\rm J}(t)
=
\sum_{i=1}^{2}(\alpha_{i}\omega_{i}^{2}/2)
\cos[\theta^{(i)}_{0}+\Lambda f(t)]
\). 
The external force terms $F_{\rm C}$ and $F_{\rm J}$ comes from,
respectively, the charge imbalance and the Cooper pair tunneling. 
The angular frequency $\omega_{\rm JL}$ is given as 
\(
 \omega_{\rm JL} = \sqrt{\alpha_{1}+\alpha_{2}}\omega_{\rm in}
\), 
which is equal to the mass of the JL mode\,\cite{Ota;Matsumoto:2009}.   
We assume $|J_{\rm in}|\gg j_{1},j_{2}$ throughout this paper. 
In this case, noting $\omega_{\rm in}^{2} \gg \omega_{i}^{2}$, 
one can safely neglect $G_{\rm J}$ in Eq.\,(\ref{eq:rp_fluctuation}). 
Then, under this assumption we have a solution of
Eq.\,(\ref{eq:rp_fluctuation}) as follows \cite{cutoff},
\begin{equation}
\widetilde{\psi}(t)
=
A\sin\Omega_{\rm mw} t
+
\sum_{k\in\mathbb{Z}}
B_{k}\sin(\Omega_{k}t+\theta_{0}^{(1)}), 
\label{eq:rlt_formal_sol}
\end{equation} 
where the coefficients $A$ and $B_{k}$ are given as
\begin{eqnarray*}
&&
A 
= 
-\xi\frac{\alpha_{1}+\alpha_{2}}{2} 
\frac{u \Omega_{\rm mw}^{2}/\omega_{\rm JL}^{2}}
{1-\Omega_{\rm mw}^{2}/\omega_{\rm JL}^{2}}, \\
&&
B_{k} 
= -\nu
\frac{j_{1}+j_{2}}{2J_{\rm in}}
\frac{J_{k}(\Lambda u)}
{1-\Omega_{k}^{2}/\omega_{\rm JL}^{2}},
\end{eqnarray*} 
with $\Omega_{k} = \Lambda \Omega_{0}+k\Omega_{\rm mw}$ and
$J_{k}(\Lambda u)$ being the Bessel function of the $k$th order.   
It should be noted that $\widetilde{\psi}$ has poles at 
$\Omega_{\rm mw}^2=\omega_{\rm JL}^2$ 
and $\Omega_{k}^2=\omega_{\rm JL}^2$, that is, $\widetilde{\psi}$ is
resonantly  enhanced at these frequencies. 
These resonant conditions can be satisfied by tuning the frequency of
the applied microwave. 
In a region far from the resonance we have $\widetilde{\psi} \simeq 0$. 
In the region we have conventional Shapiro steps as discussed above. 
It should be also noted that Eq.\,(\ref{eq:rp_fluctuation}) has a
characteristic oscillating solution corresponding to the JL mode, which is
dropped in Eq.\,(\ref{eq:rlt_formal_sol}) because the resonance effects
are predominant. 
An enhancement of the DC tunneling current by this collective mode is
pointed out without microwave irradiation as discussed in
Ref.\,\cite{Agterberg;Janko:2002}. 

Let us now study the resonance effects due to the microwave irradiation
on the $I$-$V$ characteristics in the present system.  
It is convenient to use the dimensionless parameters, $p$ and $q$,
defined as  
\begin{equation*}
 \Lambda\Omega_{0} = p \Omega_{\rm mw},
\qquad
 \Omega_{\rm mw} = q\omega_{\rm JL}.
\end{equation*}
The frequency $\Omega_k$ is then expressed as  
\(
\Omega_{k} = (p+k)q\omega_{\rm JL}
\). 
Close to the resonance we write the coefficients $A$ and
$B_{k}$ as  
\(
A
=
-\xi[(\alpha_{1}+\alpha_{2})/2]
(u/\beta_{\rm rlt})
\) and 
\(
B_{k}
=
-\nu[(j_{1}+j_{2})/2J_{\rm in}]
[J_{k}(\Lambda u)/\beta_{\rm rlt}]
\), 
where $\beta_{\rm rlt}$ is regarded as the detuning parameter. 
Let us first examine the case of $q=1$ and 
\(
k
=
k_{\pm}
\equiv
-p \pm 1
\). 
In this case the resonance takes place at 
$\Omega_{\rm mw}=\omega_{\rm JL}$ and  $\Omega_{k}=\pm\omega_{\rm JL}$, and 
thus Eq.\,(\ref{eq:rlt_formal_sol}) is approximated as  
\(
\widetilde{\psi} 
\approx 
A\sin\omega_{\rm JL}t
+B_{k_{+}}\sin(\omega_{\rm JL}t+\theta^{(1)}_{0})
+B_{k_{-}}\sin(-\omega_{\rm JL}t+\theta^{(1)}_{0})
\). 
Then, substitution of Eqs.\,(\ref{eq:theta_one}) and (\ref{eq:theta_two}) into 
Eq.\,(\ref{eq:sc}) and use of the Fourier-Bessel expansion
\,\cite{Abramowitz;Stegun:1972} for Eq.\,(\ref{eq:sc}) leads to a 
condition that a DC component emerges in $I_{\rm s}(t)$, i.e.,   
\(
 \Lambda\Omega_{0} + (n+\ell_{+}-\ell_{-})\omega_{\rm JL} =0  
\), where $n,\ell_{\pm}\in\mathbb{Z}$. 
Hence, the voltages at which the Shapiro steps appear can be
determined by the relation,  
\begin{equation*}
 p(q=1) = -n-\ell_{+}+\ell_{-} (\in \mathbb{Z}). 
\end{equation*}
Let us present the calculated results at $q=1$. 
We choose the parameter values as $u=1$,  $j_{2}/j_{1}=0.93$,
$\alpha_{2}/\alpha_{1}=1.3$, and $\alpha=0.1$. 
Figure \ref{fig:q1} shows the Shapiro steps in the cases of (a)
$\beta_{\rm rlt}=10^{-2}$ and (b) $\beta_{\rm rlt}=10^{-3}$.  
As seen in this figure, the voltage dependence of the step height
drastically changes as one approaches the resonance condition, i.e.,
from a monotonic decrease to oscillating one. 
Furthermore, one notices in Fig.\,\ref{fig:q1}(b) that in the
oscillating region a larger DC current appears at the voltages
corresponding to even (odd) $p$ in the case of the $s$-wave 
($\pm s$-wave),  that is, the Shapiro steps generated by the resonance with the JL
mode clearly depends on the pairing symmetry. 
To understand the origin of this remarkable feature in the Shapiro steps
we investigate the induced DC current for large $p$. 
It is easy to show that 
\(
\widetilde{\psi}
\approx
A \sin\omega_{\rm JL}t
\) for $|p|\gg 1$ \,\cite{large_p}. 
Then, the step height in this limit is obtained as   
\(
I_{\rm s,DC}
=
\left|
j_{1}J_{p}(x_{1})
+
j_{2}\cos\psi_{0}J_{p}(x_{2})
\right|
\), where 
\(
x_{1}
=\Lambda u + (\alpha_{1}A)/(\alpha_{1}+\alpha_{2})
\)
and 
\(
x_{2}
=\Lambda u - (\alpha_{2}A)/(\alpha_{1}+\alpha_{2})
\). 
Since $A$ is large near the resonance frequency, one may have an
approximate relation, $x_{1}\simeq -x_{2}$.  
In this case, employing the relation
$J_{n}(-x)=(-1)^{n}J_{n}(x)$ ($n\in\mathbb{Z}$ and
$x\in\mathbb{R}$), we find that 
\begin{equation*}
I_{\rm s,DC}
\approx
\left|
j_{1}+ (-1)^{p}j_{2}\cos\psi_{0}
\right|
\left|
J_{p}(x_{1})
\right|. 
\end{equation*} 
The coefficient $|j_{1}+ (-1)^{p}j_{2}\cos\psi_{0}|$ takes alternating values 
with respect to $p$, that is, 
\(
j_{1}+ (-1)^{p}j_{2}
\) 
in the $s$-wave case ($\psi_{0}=0$),  whereas 
\(
j_{1}+(-1)^{p+1}j_{2}
\)
in the  $\pm s$-wave case ($\psi_{0}=\pi$ ), which leads to the fact that 
a larger DC current appears at even (odd) $p$ in the $s$-wave ($\pm s$-wave) case.   
This remarkable feature of the Shapiro steps caused by the resonance with 
the JL mode is quite clear-cut for the determination of the pairing 
symmetry in multi-gap superconductors. 

Finally, we examine the case of $q>1$ ($q\in\mathbb{Z}$). The resonance in
this case arises from the poles in the second term on the right hand
side of Eq.\,(\ref{eq:rlt_formal_sol}), i.e.,  
$\Omega_{k}=\omega_{\rm JL}$, which brings about the correction to
$\theta^{(i)}(t)$ as 
\(
\widetilde{\psi}
\approx
B_{k}\sin(\omega_{\rm JL}t + \theta^{(1)}_{0})
\). 
Then, after similar mathematical manipulation to that in the $q=1$ case
one finds the relation, 
\(
\Lambda\Omega_{0} + n\Omega_{\rm mw} + \ell_{k}\omega_{\rm JL}
=0
\) 
with $n,\ell_{k}\in\mathbb{Z}$, for the DC current to be induced. 
From this result it follows,  
\begin{equation*}
p(\Omega_{k}=\omega_{\rm JL},q>1)
=
-n-\frac{\ell_{k}}{q}
=
-k+\frac{1}{q}. 
\end{equation*}
We note that $p$ is generally not an integer, which indicates that the
Shapiro steps appear at fractional voltages, that is, the resonance with
the JL mode induces the fractional Shapiro steps, as well. 
Figure \ref{fig:qfrac} shows the DC current, i.e., the step heights at
the resonance voltages.  
The microwave amplitude and the junction parameters are the same as in
Fig.\,\ref{fig:q1}. 
Figures \ref{fig:qfrac}(a) and (b) show the cases of $q=2$ 
(i.e., 
\(
p=m+\frac{1}{2}
\) ($m\in\mathbb{Z}$)) and $q=3$ 
(i.e., 
\(
p=m+\frac{1}{3}
\) ($m\in\mathbb{Z}$)
), respectively. 
The numerical results given in these figures proves the existence of the fractional
Shapiro steps. 
Although fractional order steps have been studied in Josephson
arrays\,\cite{Yu;Stroud:1992} and highly transmissive
junctions\,\cite{Duprat;Yeyati:2005} with conventional one-gap
superconductors, the origin in the present system is an intrinsic
feature of multi-gap superconductors, the JL modes. 

In conclusion, we revealed that an external microwave applied to the
hetero Josephson junction resonantly excites the JL mode and induces DC
currents. 
The voltage dependence of the Shapiro steps is sensitive to the
inter-band sign change, i.e., the pairing symmetry of the multi-based
superconductor.  
It is also predicted that the fractional Shapiro steps appear. 
We suggest that the tuning of the microwave frequency to the JL mode one
bring about not only unconventional types of Shapiro steps but also a clear
probe to the gap symmetry of a multi-gap superconductor.

The authors wish to acknowledge valuable discussions with
H. Matsumoto, S. Shamoto, N. Hayashi, M. Okumura, H. Nakamura, N. Nakai,
R. Igarashi, Y. Nagai, and A. Yamamoto. 
TK was partially supported by Grant-in-Aid for Scientific Research (C)
(No. 22540358) from the Ministry of Education, Culture, Sports, Science
and Technology of Japan.


\begin{thebibliography}{99}
\bibitem{Kamihara;Hosono:2008}
Y. Kamihara, T. Watanabe, M. Hirano, and H. Hosono, 
J. Am. Chem. Soc. {\bf 130}, 3296 (2008). 
\bibitem{Rotter;Johrendt:2008}
M. Rotter, M. Tegel, and D. Johrendt, 
Phys. Rev. Lett. {\bf 101}, 107006 (2008). 
\bibitem{Mazin;Du:2008}
I. I. Mazin, D. J. Singh, M. D. Johannes, and M. H. Du, 
Phys. Rev. Lett. {\bf 101}, 057003 (2008).
\bibitem{Kuroki;Aoki:2008}
K. Kuroki, S. Onari, R. Arita, H. Usui, Y. Tanaka, H. Kontani, and
	H. Aoki, 
%K. Kuroki {\it et al.}, 
Phys. Rev. Lett. {\bf 101}, 087004 (2008); 
{\bf 102}, 109902(E) (2009); 
K. Kuroki, H. Usui, S. Onari, R. Arita, and H. Aoki, 
%K. Kuroki {\it et al.}, 
Phys. Rev. B {\bf 79}, 224511 (2009). 
\bibitem{Hanaguri;Takagi:2010}
T. Hanaguri, S. Niitaka, K. Kuroki, and H. Takagi, 
Science {\bf 328}, 474 (2010). 
\bibitem{Chen;Zha:2010}
C.-T. Chen, C. C. Tsuei, M. B. Ketchen, Z.-A. Ren, and Z. X. Zha, 
%C.-T. Chen {\it et al.}, 
Nature Physics {\bf 6}, 260 (2010). 
\bibitem{Li;Zhang:2010}
%Y. Li, J. Tong, Q. Tao, C. Feng, G. Cao, Z. Xu, W. Chen, F.-C. Zhang, 
Y. Li {\it et al.}, 
New J. Phys. {\bf 12} 083008 (2010). 
\bibitem{Zhang;Takeuchi:2009}
%X. Zhang, Y. S. Oh, Y. Liu, L. Yan, K. H. Kim, R. L. Greene, and
%	I. Takeuchi, 
X. Zhang {\it et al.}, 
Phys. Rev. Lett. {\bf 102}, 147002 (2009); 
%X. Zhang, S. R. Saha, N. P. Butch, K. Kirshenbaum, J. Paglione,
%	R. L. Greene, Y. Liu, L. Yan, Y. S. Oh, K. H. Kim, and
%	I. Takeuchi, 
X. Zhang {\it et al.}, 
Appl. Phys. Lett. {\bf 95}, 062510 (2009).
\bibitem{Wu;Wu:2010}
%C. T. Wu, H. H. Chang, J. Y. Luo, T. J. Chen, F. C. Hsu, T. K. Chen,
%	M. J. Wang, and M. K. Wu, 
C. T. Wu {\it et al.}, 
Appl. Phys. Lett. {\bf 96}, 122506 (2010).
\bibitem{Kashiwaya;Kashiwaya:2010}
%H. Kashiwaya, K. Shirai, T. Matsumoto, H. Shibata, H. Kambara,
%	M. Ishikado, H. Eisaki, A. Iyo, S. Shamoto, I. Kurosawa, and
%	S. Kashiwaya, 
H. Kashiwaya {\it et al.}, 
Appl. Phys. Lett. {\bf 96}, 202504 (2010). 
\bibitem{Inotani;Ohashi:2009}
D. Inotani and Y. Ohashi,
Phys. Rev. B {\bf 79}, 224527 (2009). 
\bibitem{Linder;Sudbo:2009}
J. Linder, I. B. Sperstad, and A. Sudb\o, 
Phys. Rev. B {\bf 80}, 020503(R) (2009);  
I. B. Sperstad, J. Linder, and A. Sudb\o, 
{\it ibid}. {\bf 80}, 144507 (2009). 
\bibitem{Parker;Mazin:2009}
D. Parker and I. I. Mazin, 
Phys. Rev. Lett. {\bf 102}, 227007 (2009).
\bibitem{Wu;Phillips:2009}
J. S. Wu and P. Phillips, 
Phys. Rev. B {\bf 79}, 092502 (2009). 
\bibitem{Ng;Nagaosa:2009}
T. K. Ng and N. Nagaosa, 
EPL {\bf 87}, 17003 (2009). 
\bibitem{Golubov;Dolgov:2009}
A. A. Golubov, A. Brinkman, Y. Tanaka, I. I. Mazin, and O. V. Dolgov, 
Phys. Rev. Lett. {\bf 103}, 077003 (2009). 
\bibitem{Ota;Matsumoto:2009}
Y. Ota, M. Machida, T. Koyama, and H. Matsumoto, 
Phys. Rev. Lett. {\bf 102}, 237003 (2009); 
Phys. Rev. B {\bf 81}, 014502 (2010); 
Y. Ota, N. Nakai, H. Nakamura, M. Machida, D. Inotani, Y. Ohashi,
	T. Koyama, and H. Matsumoto, 
{\it ibid.} {\bf 81}, 214511 (2010).
\bibitem{Leggett:1966}
A. J. Leggett, 
Prog. Theor. Phys. {\bf 36}, 901 (1966). 
\bibitem{Hinken:1991}
J. H. Hinken, 
{\it Superconductor Electronics: Fundamentals and Microwave Applications} 
(Springer-Verlag, Berlin, 1991), Chap.3. 
\bibitem{cutoff}
We can find that a proper cut-off $k_{\rm c}$ exists from  
\(
|J_{k}(x)| 
\le 
\min\{
1/\sqrt{2},|x|^{k}/2^{k}\Gamma(k+1)\}
\) for $k\ge 1$ and $x\in\mathbb{R}$. 
Then, we can omit the terms for $|k|>k_{\rm c}$ in
	Eq.\,(\ref{eq:rlt_formal_sol}). 
\bibitem{Agterberg;Janko:2002}
D. F. Agterberg, E. Demler, and B. Janko, 
Phys. Rev. B {\bf 66}, 214507 (2002). 
\bibitem{Abramowitz;Stegun:1972}
{\it Handbook of Mathematical Functions with Formulas, Graphs, and
	Mathematical Tables}, 
edited by M. Abramowitz and I. A. Stegun 
(Dover, New York, 1972), Chap.9. 
\bibitem{large_p}
When $|p|>k_{\rm c}+1$, the contributions from the 3rd term in
	Eq.\,(\ref{eq:rlt_formal_sol}) to the step heights are 
	negligible. 
\bibitem{Yu;Stroud:1992}
W. Yu, E. B. Harris, S. E. Hebboul, J. C. Garland, and D. Stroud, 
Phys. Rev. B {\bf 45}, 12624 (1992). 
\bibitem{Duprat;Yeyati:2005}
R. Duprat and A. L. Yeyati, 
Phys. Rev. B {\bf 71}, 054510 (2005).  
\end{thebibliography}
\end{document}